# Towards Design and Implementation of Space Efficient and Secured Transmission scheme on E-Governance data

Mr. Nikhilesh Barik[1]   Dr. Sunil Karforma[2]   Dr. J K Mondal[3]   Ms. Arpita Ghosh[4]

1. nikhileshbarik@gmail.com   2. dr.sunilkarforma@gmail.com   3. jkm.cse@gmail.com   4. arpita_hit06@yahoo.com

**Abstract**

*We know that large amount of data and information should be transmitted through internet during transactions in E-Governance. Smart E-Governance system should deliver speedy, space efficient, cost effective and secure services among other governments and its citizens utilizing benefits of Information and Communication Technologies (ICT).*

*This paper proposes to develop a space efficient and secured data transmission scheme using Modified Huffman algorithm for compression ,which will also yield better bandwidth utilization and inner encryption technique with one way hash function SHA (Secured Hash Algorithm) to ensure Message integrity.*

**1 Introduction:**

Along with the development of Internet, the E-governance [8] has become a new pattern of activity for any Government. In the E-governance different kinds of electronic documents are processed and transmitted through communicating network. At present, the authentication of electronic documents and its efficient packaging are the main restrictions for the E-governance development. There are also some other pending problems like protection validity, uniqueness traceability of e-documents, prohibition of illegal duplication & tamper proofing. Also as in this age of universal electronic connectivity, of viruses and hackers, of electronic eavesdropping and electronic fraud, there is indeed no time at which security does not matter. Two trends have come together to make the topic of vital interest. First, the explosive growth in the usage of computer systems and their interconnections via internet has increased the dependence of organizations as well as individuals on the information stored and communicated using these systems. This in turn, has led to a heightened awareness of the need to protect data and resources from disclosure, to guarantee the authenticity of data and messages. Secondly, The Compression algorithms reduce the redundancy in data representation to decrease the storage required for that data. Data compression offers an attractive approach to reducing communication costs by using available bandwidth effectively.

The National E-Governance Plan [8] of Indian Government seeks to lay the foundation and provide for long-term growth of E-Governance within and outside the country impetuously. It's aim is to create the right governance mechanisms, set up the core infrastructure and policies to implement a number of Mission Mode Projects at the center, state to G2G concern. There are various Projects that have been taken at central service levels. Some of these are Income Tax filling ,banking services, provident fund status ,passport & visa information ,voter id / National citizen card status ,insurance & risk management ,pension plan status , details ,members of legislative ,assembly and parliament ,trade license key or agreement ,pan card verification.

In E-Governance large amount of packets would be transferred between G2G using internet which is unsecured and time consuming .Also attackers can change the base information according to their requirement .So to ensure speedy communication and reduce the unauthorized access in information and communication technologies(ICT), it is required to use some compression techniques and secured encryption techniques while data integrity is maintained using hash function so that network bandwidth will be better utilized.

One of the most important classes of cryptographic algorithms[1] in current use is the cryptographic hash functions. Hash functions are ubiquitous in today's IT systems and have a wide range of applications in security protocols and schemes, such as providing software integrity, digital signatures, message authentication and password protection. Among their many security requirements, cryptographic hash function algorithms need to feature a property known as collision resistance, that is, it must be infeasible to construct two distinct inputs with the same hash output.

Section 2 describes Framework for secured G2G model . In section 3, the techniques of the scheme are implemented on a sample E-governance database from sender Government to Receiver Government Section 4 represents analyses the proposed technique from different perspective . Section 5 draws a conclusion followed by reverences and Appendix.1.0 for output.

**2. Frame work of secured G2G Model :**

G2G model implies Government to Government[9] e-commerce activities. It can also be referred to E-commerce activities between two or more governments, central to its states/provinces or one country to other. It can be used to vertical and horizontal Governmental integration.

So G2G is the online non-commercial interaction between Government organizations, departments, and authorities and other Government organizations, departments, and authorities. G2G systems generally come in one of the two types i> Internal facing - joining up a single Government's departments, agencies, organizations and authorities and ii>External facing - joining up multiple Government's information systems.

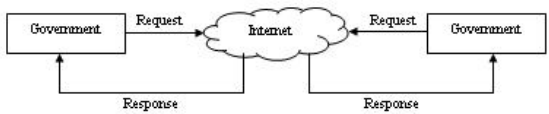

Figure 2.1        Working Model of G2G

Following block diagram is shows the detail applications of our proposed scheme to any G2G model .

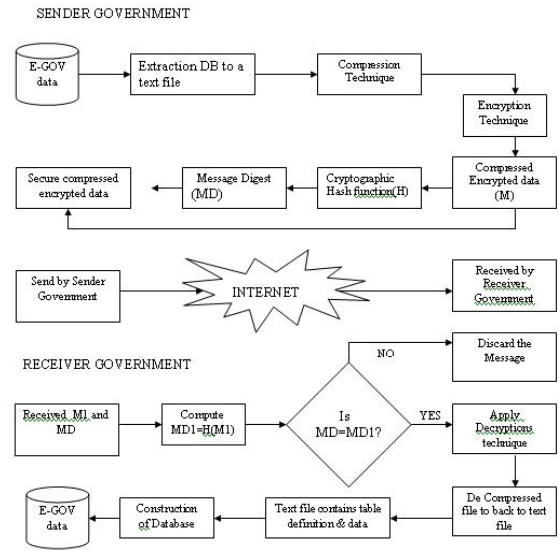

**Figure 2.2 : Block Diagram to G2G Model.**

**3. Proposed Space Efficient and Secured Transmission scheme :** The following flow diagram is describe our proposed space efficient and secured transmission scheme from sender Government to receiver Government. The proposed technique are to perform a compression[4] & encryption[2] on extracted text from sender and then to apply hash function to maintain the integrity of the data transmitted by sender Government. Receiver government first checks message digest by using hash function [1,3] and start decryption, decompression to bring back to the original database. Each part is being discussed separately along with relevant examples.

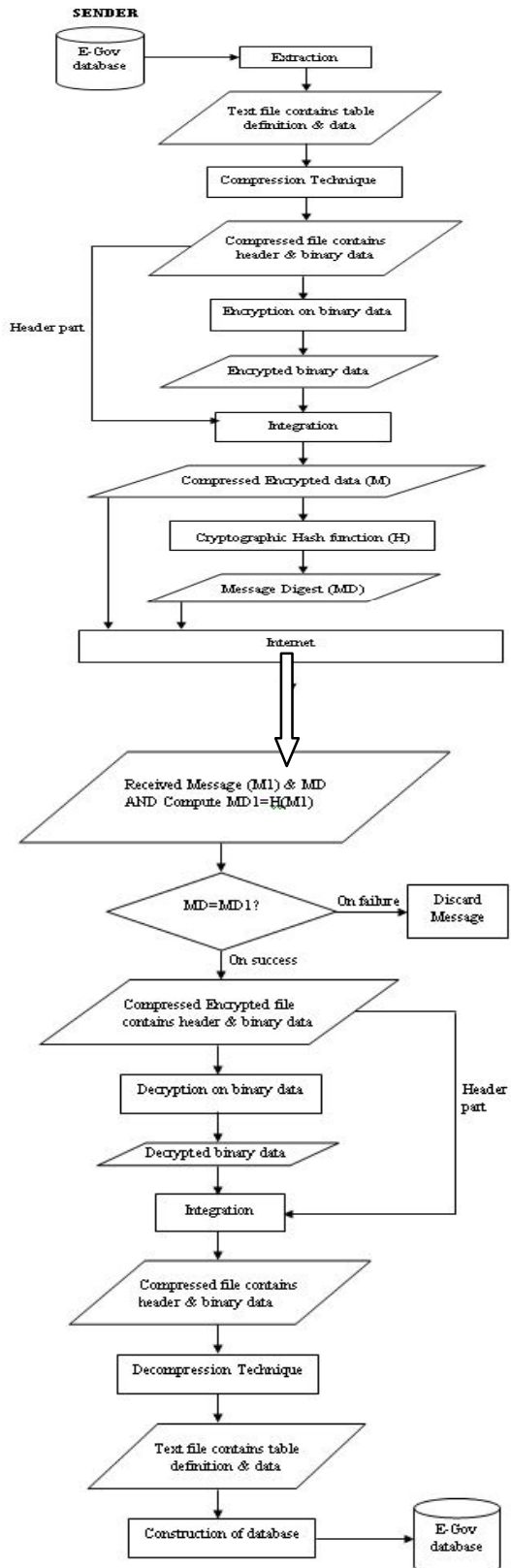

Figure:3.1 FLOW DIAGRAM OF THE SCHEME

**3.1 Compression:** At the sender a text file is being generated from the original database. Modified Huffman Algorithm has been used to compress the text file.

**3.1.1 Using modified Huffman Algorithm:**

**Step1.** The extracted text file is taken as input. Individual character and their probabilities of occurrences are determined.

**Step2.** Probability of occurrences of each of the characters is considered as the weight of that character.

**Step3.** Huffman tree is constructed from the sorted weights considering the minimum weights as the leaf node.

**Step4.** Visiting all of the external nodes starting from the root node every time generates binary encoded string corresponding to individual character.

**Step5.** Shorter bit strings represent the characters in the input text file with more occurrences and longer bit strings represent characters with fewer occurrences.

**Step6.** The compressed file having individual character, length of its corresponding binary encoded string and the binary encoded string (together known as header) is created .

**Step7.** The binary encoded string corresponding to the content of the input text file in compressed form is stored after the header part.

The primary difficulty associated with variable-length code words is that the rate at which bits are presented to the transmission channel will fluctuate, depending on the relative frequencies of the character present in the source messages. This requires buffering between the source and the channel. Advances in technology have done both overcame this difficulty and contributed to the appeal of variable-length codes.

Current data networks allocate communication resources to sources on the basis of need and provide buffering as part of the system. These systems require significant amounts of protocol, and fixed-length codes are quite inefficient for applications such as packet headers. In addition, communication costs are beginning to dominate storage and processing costs, so that variable-length coding schemes which reduce communication costs are attractive even if they are more complex. For these reasons, one could expect to see even greater use of variable-length coding in the future.[4]

**3.2 The Encryption at Sender end [7]**: A stream of bits is considered as the plaintext. The plaintext is divided into a finite number of blocks, each having a finite number of bits. The proposed technique is then applied to encrypt the data in unit of blocks.

```
Evaluate: D_L the decimal equivalent, corresponding to the
source block S = s_0 s_1 s_2 s_3 s_4 ... s_{L-1}.
Set: P = 0.
LOOP: Evaluate: Temp = Remainder of D_{L,P} / 2.
        If Temp = 0
                Evaluate: D_{L,P+1} = D_{L,P} / 2.
                Set: t_P = 0.
        Else If Temp = 1
                Evaluate: D_{L,P+1} = (D_{L,P} + 1) / 2.
                Set: t_P = 1.
        Set: P = P + 1.
        If (P > (L – 1)) Exit.
ENDLOOP
```

**Figure 3.2: Pseudocode to Encrypt a Source Block wise.**

**3.3 Use cryptographic Hash SHA to maintain data integrity :** Cryptographic hash function[1] will be used to verify the integrity of the data i.e. to ensure that the message has not been tempered with after it leaves the sender but before it reaches the receiver. So we perform the hash operations (some time called message digest algorithm) over a block of decompressed and encrypted data to produce its hash, which is smaller in size than the original message[5,6].

In general hash function H(M), which operates on our compressed and encrypted message M. It returns a fixed-length hash value MD.

So that MD = H(M),

Given M, it is easy to compute MD

Given MD, it is hard to compute M such that H(M)=MD.

Given M, it is hard to find another message, M1, such that H(M) =H(M1).

If any two messages produce the same message digest ,thus violating the principle which is known as collision. But message digest algorithm usually produce a message digest of length 128 bits or 160 bits, so the chances of any two message digests being the same are one in $2^{128}$ or $2^{160}$ respectively, which is possible in theory but rare in practice.

If a message digest uses 64 bit key then after trying $2^{64}$ transaction ,an attacker can expect that for two different messages we may get the same message digest.

Among the different One-Way Hash Functions like Snefru, N-Hash,MD4,MD5,MD2, SHA (Secure Hash Algorithm),RIPE-MD,HAVAL we are going to apply SHA-512 which takes a message of $2^{128}$ bit and produce a message digest of 512 bits.

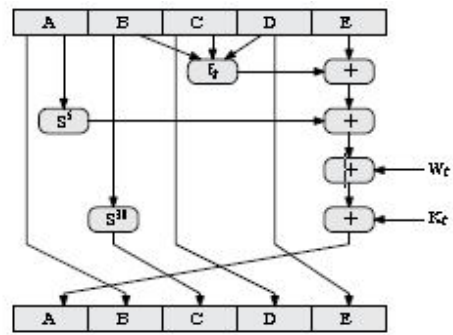

**Figure:3.3 Elementary SHA Operation (single step) [3]**

**3.4 Decryption from Receiver end :** Receiver Government now convert the encrypted message into the corresponding stream of bits and then this stream is to be decomposed into a finite set of blocks, each consisting of a finite set of bits. Now, during this process of decomposition, the way by which the source stream was decomposed during encryption is to be followed, so that corresponding to each block, which is effectively the target block, the source block can be generated. We discuss that scheme under the proposed technique, through which a source block is generated from a block in the encrypted stream of bits.

```
Set: P = L – 1 and T = 1.
LOOP:
    If t_P = 0
        Evaluate: T = T^th even number in the series
        of natural numbers.
    Else
        If t_P = 1
            Evaluate: T = T^th odd number in the series of
            natural numbers.
    Set: P = P – 1.
    If P < 0
        Exit.
ENDLOOP
Evaluate: S = s0 s1 s2 s3 s4 … sL-1, which is the
binary equivalent of T.
```

**Figure 3.4 : Pseudocode to decrypt a target text Block wise.**

### 3.5 Decompression from Receiver end

Now receiver Government transfer the binary format to text file using header. De compression algorithm used to decompressed. Note that the last character will be retrieved according to the padding value is written very after $$(indicating end of header).

### 3.6 Import this text file to required database.

Ultimate objective is to get the required database which was send from the sender Government.

**No alteration in data in the database.**

If it is possible to import this text file as a database then we come to conclusion that no alteration of data in table of the concern database.

**Efficiency of techniques.**

As in total at most sixty different characters (including all numeric , alphanumeric and special characters ) can be used so header will be fixed for any number of fields and records. Hence the significance of using compression algorithm is justified for however large the database of E-Governance data may be.

| Original Size(byte) | 1190 | 2384 | 8336 | 11177 | 22358 | 37266 | 81990 |
|---|---|---|---|---|---|---|---|
| Compressed Size(byte) | 935 | 1772 | 6022 | 7900 | 15295 | 25155 | 53374 |
| Percentage | 78.57 | 74.33 | 72.24 | 70.68 | 68.40 | 67.50 | 65.09 |

Figure 3.5: Compression ratio.

Hence we can neglect the size of header at the time of compression calculation .
**Sample output is shown in Appendix 1.0**
Also it is secure enough as data integrity can be checked by hash function.

### 4. Analyzing the proposed technique from different perspective:

As there is general lack of awareness regarding the benefits of E-Governance as well as the process involved in implementing successful G2G (Government to Government) projects ,the administrative structure is not geared for maintaining, storing and retrieving governance information electronically and transactions are be made in non--secured way . So day by day corruption is increasing from different ends. The privacy of each user and system data is of crucial importance for E-Governance system that meets user expectations and acceptance. Here we are proposing an authorization model accompanying basic security and space efficient mechanisms. This mechanisms can also be applied in different fields like G2B or G2C for secure data transmission scheme.

### 5. Conclusion:

In this paper, we have proposed to use Modified Huffman Algorithm for data compression as well as better bandwidth utilization of internet .Also we used encryption and secured hash algorithm to maintain data integrity by SHA hash function through which E-Governance transaction take place. As E-Governance can be described as the internal Government Operations as well as mediator between the Government and Citizen or Business to enhance the efficiency of the aspects of Government. The result will be much more efficient when we consider a large database as well as use of strong compression and encrypted algorithm and strong one way hash function.

### 6. References:

1. Schneier Bruce, "Applied Cryptography, Second Edition: Protocols, Algorthms, and

1. Source Code in C (cloth), Publisher: John Wiley & Sons, Inc.
2. Stallings Williams, Cryptography and Network Security – Principal and Practices, Pearson Education
3. Kahate Atul "CRYPTOGRAPHY and NETWORK SECURITY, Second Edition,
4. Samanta ,D, "Classic Data Stucture", PHI Publications
5. Karforma Sunil and Banerjee Sanjay "A prototype design for DRM based credit card transaction in E-commerce" Pub: ACM New York, NY, USA
6. Karforma Sunil and Mukhopadhyay Sripati "A Secured Banking Transaction System using Digital Signature Algorithm"
7. Dutta S. and Mandal J. K., "Ensuring E-Security Using A Private-Key Cryptographic System Following Recursive Positional Modulo-2 Substitutions.
8. http://india.gov.in/govt/national_egov_plan.php
9. http://en.wikipedia.org/wiki/Government_to_Government
10. http://www.shadowtech-asp.net/examples/crypto

**Appecndix 1.0**

We consider a small database as E_GOV and a table under that SSN having following three fields

| SSN_ID | Bigint(10) |
|---|---|
| PASSPORT_CODE | Varchar(20) |
| MOBILE | Bigint(10) |

Table :SSN

| SSN_ID | PASSPORT_CODE | MOBILE |
|---|---|---|
| WB191134355525 | DAIBIKJ33998822 | 9434538808 |
| WB191134385585 | NBARIK12349876 | 9434516929 |

**Figure A1: Sample data structure of Table SSN Extracted a database to a text file**

Corresponding text file "E-GOV.txt"of the above database can be exported as:

===Database E-GOV

== Table structure for table ssn

|------
|Field|Type|Null|Default
|------
|SSN_ID|bigint(10)|No|
|PASSPORT_CODE|varchar(20)|No|
|MOBILE|bigint(10)|No|
== Dumping data for table ssn
| WB191134355525|DAIBIKJ33998822|9434538808
| WB191134385585|NBARIK12349876|9434516929

A stream of bits is considered as the plaintext. The plaintext is divided into a finite number of blocks, each having a finite number of bytes. The algorithm is then applied for each of the blocks. Considering one block 8-bits of source block is 01100011  to be encrypted. So encrypted binary data corresponding the code will be 10111001 according to the Algorithm in the figure 3.2

**Compressed and encrypted code(M).**
c╠ Ö-¯ õõ7œáe™Dc6N¿}7õ¹º5®æ›+wm~úoï²D¯ ffh!‹£ã ,g]P{sÞ:¡Z¹ö@fff,ª½f   Cÿ
ó+8Ù3☼ p▯   ×ª§rÖ&‹½g‼ šæ4ö¹qRf-à╠ LJF3yÇþ╢ æV q²f-à◄ ŒßWO°ó"÷-
VÖÊÃ  Û_¾ûì'╚ Z:¯‰↑ Àšã…¯¾Ô »ø¾/   ÷'h→ ¡é╚ ëc
 ªâõ÷
╠ .õq|VÄZK

**Application of Hash on message M:** For above M ,message digest(using SHA1 algorithm) [10] MD will be

*─ ◆-#◆ec◆◆L◆K◆g← ◆◆u╚ 1

which also will be send from the sender including with the message(M).

**Decryption algorithm applied :**After successful verification by message digest of sender data, the decryption technique applied    with the target block 10111001 is decrypted as "01100011" according to the decryption algorithm in figure 3.4 and it will be applicable for each block.

**Decompression algorithm applied:** Decompression algorithm will be applied to the decryption data from the receiver end and followed by text file to be imported to get the require database as per Figure A